\documentclass[a4paper,aps,preprintnumbers,showpacs,amsmath,amssymb,prl]{revtex4}
\usepackage{mathrsfs}
\usepackage{amsmath}
\usepackage[dvips]{graphicx}
\usepackage{epsfig}
\usepackage[dvips]{graphics,graphicx}
\begin{document}

\title{Cavity Quantum Optomechanics of Ultracold Atoms in an Optical Lattice: Normal-Mode Splitting}

\author{Aranya B Bhattacherjee}
\affiliation {Max Planck-Institute f\"ur Physik komplexer Systeme, N\"othnitzer Str.38, 01187 Dresden, Germany} \affiliation {Department of Physics, ARSD College, University of Delhi(Souh Campus), New Delhi-110021, India.}

\begin{abstract}
We consider the dynamics of a movable mirror (cantilever) of a cavity coupled through radiation pressure to the light scattered from ultracold atoms in an optical lattice. Scattering from different atomic quantum states creates different quantum states of the scattered light, which can be distinguished by measurements of the displacement spectrum of the cantilever. We show that for large pump intensities the steady state displacement of the cantilever shows bistable behaviour. Due to atomic back-action, the displacement spectrum of the cantilever is modified and depends on the position of the condensate in the Brillouin zone. We further analyze the occurrence of splitting of the normal mode into three modes due to mixing of the mechanical motion with the fluctuations of the cavity field and the fluctuations of the condensate with finite atomic two-body interaction. The present system offers a novel scheme to coherently control ultracold atoms as well as cantilever dynamics.
\end{abstract}


\maketitle

\section{Introduction}

In recent years two distinct subjects, optical-micro cavities and nano-mechanical resonators have become entangled experimentally by underlying mechanism of optical, radiation pressure forces. The coupling of mechanical and optical degrees of freedom via radiation pressure has been a subject of early research in the context of laser cooling \cite{hansch,wineland, chu} and gravitational-wave detectors \cite{caves}. Recently there has been a great surge of interest in the application of radiation forces to manipulate the center-of-mass motion of mechanical oscillators covering a huge range of scales from macroscopic mirrors in the Laser Interferometer Gravitational Wave Observatory (LIGO) project \cite{corbitt1, corbitt2} to nano-mechanical cantilevers\cite{hohberger, gigan, arcizet, kleckner, favero, regal}, vibrating microtoroids\cite{carmon, schliesser} membranes\cite{thompson} and Bose-Einstein condensates \cite{brennecke, murch}. The quantum optical properties of a mirror coupled via radiation pressure to a cavity field show interesting similarities to an intracavity Kerr-like interaction \cite{fabre}. Recently, in the context of classical investigations of nonlinear regimes, the dynamical instability of a driven cavity having a movable mirror has been investigated \cite{marquardt}. Theoretical work has proposed to use the radiation-pressure coupling for quantum non-demolition measurements of the light field \cite{braginsky}.

In the field of quantum degenerate gases, standard methods to observe quantum properties of ultracold atoms are based on destructive matter-wave interference between atoms released from traps \cite{Greiner}. Recently, a new approach was proposed which is based on all optical measurements that conserve the number of atoms. It was shown that atomic quantum statistics can be mapped on transmission spectra of high-Q cavities, where atoms create a quantum refractive index. This was shown to be useful for studying phase transitions between Mott insulator and superfluid states since various phases show qualitatively distinct spectra \cite{Mekhov07}. New possibilities for cavity opto-mechanics by combining the tools of cavity electrodynamics with those of ultracold gases is the motivation of the present work. Here we show that different quantum states of ultracold gases in optical lattice, confined in a cavity  can be distinguished by the steady state displacement spectrum of the movable mirror (cantilever). The atomic and cantilever back-action shifts the cavity resonance. The laser-pump is shown to coherently control the dynamics of the mirror. Changing the pump intensity, one can switch between stable and bistable regimes. Due to coupling between the condensate wavefunction and the cantilever, mediated by the cavity photons, the cantilever displacement spectrum is continuously modified as the condensate moves across the Brillouin zone. We also show that in the presence of atom-atom interactions, the coupling of the mechanical oscillator, the cavity field fluctuations and the condensate fluctuations (Bogoliubov mode) leads to the splitting of the normal mode into three modes (Normal Mode Splitting).

\section{Cantilever displacement spectra as a probe of quantum phases of ultracold atoms}

\begin{figure}[t]
\hspace{1.0cm}
\includegraphics [scale=0.5] {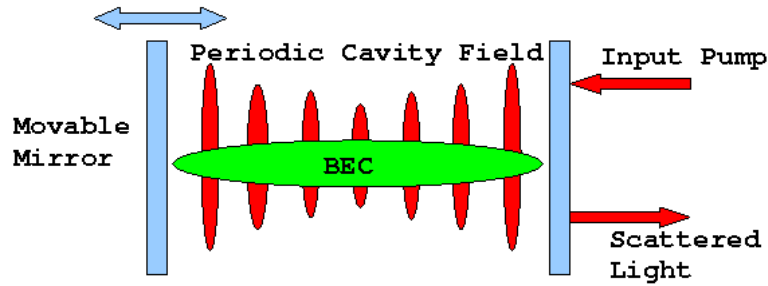} 
\caption{Optomechanical realization of parametric coupling of a mechanical oscillator to a cavity optical mode and Bose-Einstein condensate trapped in an optical lattice inside the cavity.}
\label{figure1}
\end{figure}

We consider an elongated cigar shaped Bose-Einstein condensate(BEC) of $N$ two-level $^{87} Rb$ atoms in the $|F=1>$ state with mass $m$ and frequency $\omega_{a}$ of the $|F=1>\rightarrow |F'=2>$ transition of the $D_{2}$ line of $^{87} Rb$, strongly interacting with a quantized single standing wave cavity mode of frequency $\omega_{c}$ (Fig.1). The cavity field is also coupled to external fields incident from one side mirror.   In order to create an elongated BEC, the frequency of the harmonic trap along the transverse direction should be much larger than one in the axial (along the direction of the optical lattice) direction. The system is also coherently driven by a laser field with frequency $\omega_{p}$ through the cavity mirror with amplitude $\eta$. The cavity mode is coupled to a mechanical oscillator (movable mirror) with frequency $\Omega_{m}$ via the dimensionless parameter $\epsilon=(x_{0}/\Omega_{m})\dfrac{d \omega_{p}}{dx}|_{x=0}$. Here $x_{0}=\sqrt{\hbar/2m_{eff}\Omega_{m}}$ is the zero point motion of the mechanical mode, $m_{eff}$ is its effective mass. It is well known that high-Q optical cavities can significantly isolate the system from its environment, thus strongly reducing decoherence and ensuring that the light field remains quantum-mechanical for the duration of the experiment. The harmonic confinement along the directions perpendicular to the optical lattice is taken to be large so that the system effectively reduces to one-dimension. This system is modeled by the opto-mechanical Hamiltonian $(H_{om})$ in a rotating wave and dipole approximation.

\begin{eqnarray}\label{hom}
H_{om}&=&\dfrac{p^2}{2m}-\hbar \Delta_{a} \sigma^{+} \sigma^{-} -\hbar \Delta_{c}\hat{a}^{\dagger}\hat{a}+\hbar \Omega_{m} \hat{a_{m}}^{\dagger}\hat{a_{m}} \nonumber \\&-&i\hbar g(x)\left[ \sigma^{+}\hat{a}-\sigma^{-}\hat{a}^{\dagger}\right]-i\eta(\hat{a}-\hat{a}^{\dagger})+\hbar \epsilon \Omega_{m} \hat{a}^{\dagger}\hat{a}(\hat{a}_{m}+\hat{a}_{m}^{\dagger})\,
\end{eqnarray}

where $\Delta_{a}=\omega_{p}-\omega_{a}$ and $\Delta_{c}=\omega_{p}-\omega_{c}$ are the large atom-pump and cavity-pump detuning, respectively.  Here $\sigma^{+} , \sigma^{-}$ are the Pauli matrices. The atom-field coupling is written as $g(x)=g_{0} \cos(kx)$. Here $\hat{a}$ and $\hat{a}_{m}$ are the annihilation operators for a cavity photon and the mechanical mode respectively. The input laser field populates the intracavity mode which couples to the cantilever through the radiation pressure and the atoms through the dipole interaction. The field in turn is modified by the back-action of the atoms and cantilever. It is important to notice the nonlinearity in Eqn. (\ref{hom}) arising from the coupling between the intracavity intensity and the position quadrature of the cantilever. The system we are considering is intrinsically open as the cavity field is damped by the photon-leakage through the massive coupling mirror and the cantilever is connected to a bath at finite temperature. In the absence of the radiation-pressure coupling, the cantilever would undergo a pure Brownian motion driven by its contact with the thermal environment. Here we also assume that there is no direct coupling between the atoms and the cantilever, though this coupling could give rise to some interesting physics. Since the detuning $\Delta_{a}$ is large, spontaneous emission is negligible and we can adiabatically eliminate the excited state using the Heisenberg equation of motion $\dot{\sigma^{-}}=\dfrac{i}{\hbar}\left[ H_{om},\sigma^{-}\right] $. This yields the single particle Hamiltonian

\begin{eqnarray}
H_{0}&=&\dfrac{p^2}{2m}-\hbar \Delta_{c}\hat{a}^{\dagger}\hat{a}+\cos^2(kx)\left[ V_{\text{cl}}({\bf{r}})+\hbar U_{0}\hat{a}^{\dagger} \hat{a}\right] \nonumber \\&-&i\eta(\hat{a}-\hat{a}^{\dagger})+\hbar \Omega_{m} \hat{a_{m}}^{\dagger}\hat{a_{m}}+\hbar \epsilon \Omega_{m} \hat{a}^{\dagger}\hat{a}(\hat{a}_{m}+\hat{a}_{m}^{\dagger}).
\end{eqnarray}

The parameter $U_{0}=\dfrac{g_{0}^{2}}{\Delta_{a}}$ is the optical lattice barrier height per photon and represents the atomic backaction on the field \cite{Maschler05}. $V_{\text{cl}}({\bf{r}})$ is the external classical potential. Here we will always take $U_{0}>0$. In this case the condensate is attracted to the nodes of the light field and hence the lowest bound state is localized at these positions which leads to a reduced coupling of the condensate to the cavity compared to that for $U_{0}<0$.  Along $x$, the cavity field forms an optical lattice potential of period $\lambda/2$ and depth ($\hbar U_{0}<\hat{a}^{\dagger}\hat{a}>+V_{cl}$). We now write the Hamiltonian in a second quantized form including the two body interaction term.

\begin{eqnarray}
H&=&\int d^3 x \Psi^{\dagger}(\vec{r})H_{0}\Psi(\vec{r})\nonumber \\&+&\dfrac{1}{2}\dfrac{4\pi a_{s}\hbar^{2}}{m}\int d^3 x \Psi^{\dagger}(\vec{r})\Psi^{\dagger}(\vec{r})\Psi(\vec{r})\Psi(\vec{r})\;
\end{eqnarray}

where $\Psi(\vec{r})$ is the field operator for the atoms. Here $a_{s}$ is the two body $s$-wave scattering length. The corresponding opto-mechanical-Bose-Hubbard (OMBH) Hamiltonian can be derived by writing $\Psi(\vec{r})=\sum_{j} \hat{b}_{j} w(\vec{r}-\vec{r}_{j})$, where $w(\vec{r}-\vec{r}_{j})$ is the Wannier function and $\hat{b}_{j}$ is the corresponding annihilation operator for the bosonic atom at the $j^{th}$ site. Retaining only the lowest band with nearest neighbor interaction, we have

\begin{eqnarray}
H &=& E_{0}\sum_{j}\hat{b}_{j}^{\dagger}\hat{b}_{j}+E\sum_{j}\left(\hat{b}_{j+1}^{\dagger}\hat{b}_{j}+\hat{b}_{j+1}\hat{b}_{j}^{\dagger} \right)\nonumber \\&+& (\hbar U_{0}\hat{a}^{\dagger}\hat{a}+V_{cl})\left\lbrace J_{0}\sum_{j}\hat{b}_{j}^{\dagger}\hat{b}_{j}+J \sum_{j}\left(\hat{b}_{j+1}^{\dagger}\hat{b}_{j}+\hat{b}_{j+1}\hat{b}_{j}^{\dagger} \right)\right\rbrace+\dfrac{U}{2}\sum_{j}\hat{b}_{j}^{\dagger}\hat{b}_{j}^{\dagger}\hat{b}_{j}\hat{b}_{j}\nonumber \\&-&\hbar \Delta_{c} \hat{a}^{\dagger}\hat{a}-i\hbar \eta (\hat{a}-\hat{a}^{\dagger})+\hbar \Omega_{m} \hat{a_{m}}^{\dagger}\hat{a_{m}}+\hbar \epsilon \Omega_{m} \hat{a}^{\dagger}\hat{a}(\hat{a}_{m}+\hat{a}_{m}^{\dagger})\;
\end{eqnarray}

where

\begin{eqnarray}
U&=&\dfrac{4\pi a_{s}\hbar^{2}}{m}\int d^3 x|w(\vec{r})|^{4}\nonumber \\
E_{0}&=&\int d^3 x w(\vec{r}-\vec{r}_{j})\left\lbrace \left( -\dfrac{\hbar^2 \nabla^{2}}{2m}\right) \right\rbrace w(\vec{r}-\vec{r}_{j})\nonumber \\
E &=&\int d^3 x w(\vec{r}-\vec{r}_{j})\left\lbrace \left( -\dfrac{\hbar^2 \nabla^{2}}{2m}\right) \right\rbrace w(\vec{r}-\vec{r}_{j \pm 1})\nonumber \\
J_{0}&=&\int d^3 x w(\vec{r}-\vec{r}_{j}) \cos^2(kx)w(\vec{r}-\vec{r}_{j})\nonumber \\
J &=&\int d^3 x w(\vec{r}-\vec{r}_{j}) \cos^2(kx)w(\vec{r}-\vec{r}_{j \pm 1}).
\end{eqnarray}

The OMBH Hamiltonian derived above is valid only for weak atom-field nonlinearity \cite{larson}. The nearest neighbor nonlinear interaction terms are usually very small compared to the onsite interaction and are neglected as usual. We now write down the Heisenberg-Langevin equation of motion for the bosonic field operator $\hat{b}_{j}$ , the internal cavity mode $\hat{a}$ and the mechanical mode $\hat{a}_{m}$ as

\begin{eqnarray}
\dot{\hat{b}}_{j}&=&-i(U_{0}\hat{a}^{\dagger} \hat{a}+\dfrac{V_{cl}}{\hbar}) \left\lbrace J_{0}\hat{b}_{j}+J (\hat{b}_{j+1}+\hat{b}_{j-1}) \right\rbrace-\dfrac{iE}{\hbar}\left\lbrace\hat{b}_{j+1}+\hat{b}_{j-1}  \right\rbrace\nonumber \\&-& \dfrac{iU}{\hbar}\hat{b}^{\dagger}_{j} \hat{b}_{j} \hat{b}_{j}-\dfrac{iE_{0}}{\hbar} \hat{b}_{j}\;
\end{eqnarray}

\begin{eqnarray}\label{lngvn1}
\dot{\hat{a}}&=&-iU_{0}\left\lbrace J_{0}\sum_{j}\hat{b}_{j}^{\dagger}\hat{b}_{j}+J \sum_{j} \left(\hat{b}_{j+1}^{\dagger}\hat{b}_{j}+\hat{b}_{j+1}\hat{b}_{j}^{\dagger} \right)\right\rbrace \hat{a}+\eta\nonumber \\&+&i \left\lbrace \Delta_{c} -\epsilon \Omega_{m} (\hat{a}_{m}+\hat{a_{m}}^{\dagger}) \right\rbrace \hat{a}-\dfrac{\kappa}{2} \hat{a}+\sqrt{\kappa} \xi_{p}(t)\;
\end{eqnarray}

\begin{equation}\label{lngvn2}
\dot{\hat{a}}_{m}=(-i\Omega_{m}-\dfrac{\Gamma_{m}}{2})\hat{a}_{m}-i \epsilon \Omega_{m}\hat{a}^{\dagger}\hat{a}+\sqrt{\Gamma_{m}}\xi_{m}(t).
\end{equation}

Here $\kappa$ and $\Gamma_{m}$ characterizes the dissipation of the optical and mechanical degree of freedom resectively. Here, we follow a semi-classical theory by considering \textit{noncommuting} noise operators for the input field, i.e., $\langle\xi_{p}(t)\rangle=0$, $\langle\xi_{p}^{\dagger}(t^{\prime})\xi_{p}(t)\rangle=n_{p}\delta(t^{\prime}-t),\,\langle\xi_{p}(t^{\prime})\xi_{p}^{\dagger}(t)\rangle=\left(n_{p}+1\right)\delta(t^{\prime}-t)$, and a \textit{classical} thermal noise input for the mechanical oscillator,
i.e.~$\langle\xi_{m}(t)\rangle=0$, $\langle\xi_{m}^{\dagger}(t^{\prime})\xi_{m}(t)\rangle=\langle\xi_{m}(t^{\prime})\xi_{m}^{\dagger}(t)\rangle=n_{m}\delta(t^{\prime}-t)$, in Eqs.~(\ref{lngvn1},\ref{lngvn2}). The quantities $n_{m}$ and $n_{p}$ are the equilibrium occupation numbers for the mechanical and optical oscillators, respectively. We consider a deep lattice formed by a strong classical potential $V_{\text {cl}}({\bf r})$, so that the overlap between Wannier functions is small. Thus, we can neglect the contribution of tunneling by putting $E=0$ and $J=0$ . Under this approximation, the matter-wave dynamics is not essential for light scattering. In experiments, such a situation can be realized because the time scale of light measurements can be much faster than the time scale of atomic tunneling. One of the well-known advantages of the optical lattices is their extremely high tunability. Thus, tuning the lattice potential, tunneling can be made very slow~\cite{jaksch}. 

The steady state value of the position quadrature $x_{m,s}=\hat{a}_{m,s}+\hat{a}_{m,s}^{\dagger}$ (the subscript $s$ denotes the steady state value) is found as

\begin{equation}\label{mechanical_steady}
 x_{m,s}=\dfrac{-8\epsilon \Omega_{m}^{2} \hat{a}_{s}^{\dagger}\hat{a}_{s}}{4\Omega_{m}^{2}+\Gamma_{m}^{2}},
\end{equation}

where, 

\begin{equation}\label{optical_steady}
\hat{a}_{s}^{\dagger}\hat{a}_{s}=\dfrac{\eta^{2}}{{(\Delta_{c}-U_{0}J_{0} \hat{N}-\epsilon \Omega_{m} x_{m,s})^{2}+\kappa^{2}/4}}.
\end{equation}

\begin{figure}[t]
\includegraphics[scale=0.7]{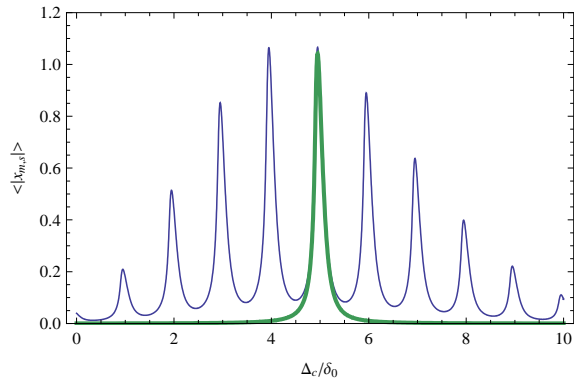}
\caption{Steady state displacement spectra of the cantilever. The single Lorentzian (thick line) for the MI reflects the non-fluctuating atom number. Many Lorentzians for SF (thin line) reflect atom number fluctuations. Here we have used $g_{m}/\delta_{0}=0.2$, $\Omega_{m}/\delta_{0}=30$, $\Gamma_{m}/\delta_{0}=0.001$ and $\kappa/\delta_{0}=0.1$}.
\label{fig.2}
\end{figure}

Here $\hat {N}=\hat {b}_{j}^{\dagger} \hat {b}_{j}$. The equation for $\hat{a}_{s}^{\dagger}\hat{a}_{s}$ is very important for the understanding the physics behind this problem. From it we clearly see how the cantilever and atom dynamics affects the steady state of the intracavity field. The coupling to the mirror and the atoms shifts the cavity resonance frequency and changes the field inside the cavity in a way to induce a new stationary intensity. The change occurs after a transient time depending on the response of the cavity and the strength of the coupling to the cantilever and the ultracold atoms.  Following \cite{Mekhov07} Eqs.~(\ref{mechanical_steady},\ref{optical_steady}) allows to express  $x_{m,s}$ as a function $f(\hat{n}_1,...,\hat{n}_M)$ of atomic occupation number operators and calculate their expectation values for prescribed atomic states $|\Psi\rangle$. 

For the Mott state  $\langle\hat{n}_j\rangle_\text{MI}=q_j$ atoms are well localized at the $j$th site with no number fluctuations. It is represented by a product of Fock states, i.e. $|\Psi\rangle_\text{MI}=\prod_{j=1}^M |q_j\rangle_j\equiv |q_1,...,q_M\rangle$, with expectation values

\begin{eqnarray}\label{14}
\langle f(\hat{n}_1,...,\hat{n}_M)\rangle_\text{MI}=f(q_1,...,q_M),
\end{eqnarray}

For simplicity we consider equal average densities $\langle\hat{n}_j\rangle_\text{MI}=N/M\equiv n_{0}$ .

In SF state, each atom is delocalized over all sites leading to local number fluctuations. It is represented by superposition of Fock states corresponding to all possible distributions of $N$ atoms at $M$ sites: $|\Psi\rangle_\text{SF} =\sum_{q_1,...,q_M}\sqrt{N!/M^N}/\sqrt{q_1!...q_M!} |q_1,...,q_M\rangle$.  Expectation values of light operators can be calculated from 

\begin{eqnarray}\label{15}
\langle f(\hat{n}_1,...,\hat{n}_M)\rangle_\text{SF}=\frac{1}{M^N}
\sum_{q_1,...,q_M}\frac{N!} {q_1!...q_M!}f(q_1,...,q_M),
\end{eqnarray}

representing a sum of all possible ``classical'' terms. Thus, all these distributions contribute to scattering from a SF, which is
obviously different from $\langle f(\hat{n}_1,...,\hat{n}_M)\rangle_\text{MI}$ with only a single contributing term.

In Fig.2, we represent $<|x_{m,s}|>$ as a function of $\Delta_{c}/\delta_{0}$, where $\delta_{0}=U_{0}J_{0}$. Clearly for the SF case, the displacement spectra is a sum of Lorentzians with different dispersion shifts. A comb like structure is seen if each Lorentzian is resolved. In the Mott state however, a single Lorentzian (thick line) is noticed. These were the typical structures predicted for the transmission spectra\cite{Mekhov07}. Since the coupling between the cantilever and the condensate wavefunction is mediated by the cavity field, we find that the displacement spectrum of the cantilever maps the distribution function of the ultracold atoms. An important condition to resolve the Lorentzians is $\kappa< \delta_{0}$ and this condition is easily met in present experiments \cite{bourdel}. Detecting the mirror's motion is straight-forward, since the optical phase shift is directly proportional to the mirror's displacement. Typically, the lorentizian frequency spectrum of the mirror's position is obtained in this way. The peak width yields the total damping rate, including the effective opto-mechanical damping. The area under the spectrum reveals the variance of the mirror displacement., which is a measure of the effective temperature.

\section{Bistable behaviour}

We now consider the case of large number of atoms and hence treat the BEC within the mean field framework and assume the tight binding approximation where we replace $\hat{b}_{j}$ by $\phi_{j}$ and look for solutions in the form of Bloch waves $\phi_{j}=u_{k}exp(ikjd)exp(-i\mu t/ \hbar)$ \cite{bhattacherjee}. Here $\mu$ is the chemical potential, $d$ is the periodicity of the lattice and $\dfrac{1}{M}\sum_{j}\hat{b}_{j}^{\dagger} \hat{b}_{j}= n_{0}$ (atomic number density), $M$ is the total number of lattice sites. From Eqns. (\ref{mechanical_steady}) and (\ref{optical_steady}), we obtain a cubic equation in $x_{m,s}$.

\begin{equation}
x_{m,s}^{3}-\dfrac{2 \Delta}{\epsilon \Omega_{m}}x_{m,s}^{2}+\dfrac{(\Delta^{2}+\kappa^{2}/4)}{\epsilon^{2} \Omega_{m}^{2}}+\dfrac{8 \eta^{2}}{\epsilon(4 \Omega_{m}^{2}+\Gamma_{m}^{2})}= 0.
\end{equation}

\begin{figure}[t]
\begin{tabular}{cc}
\includegraphics[scale=0.7]{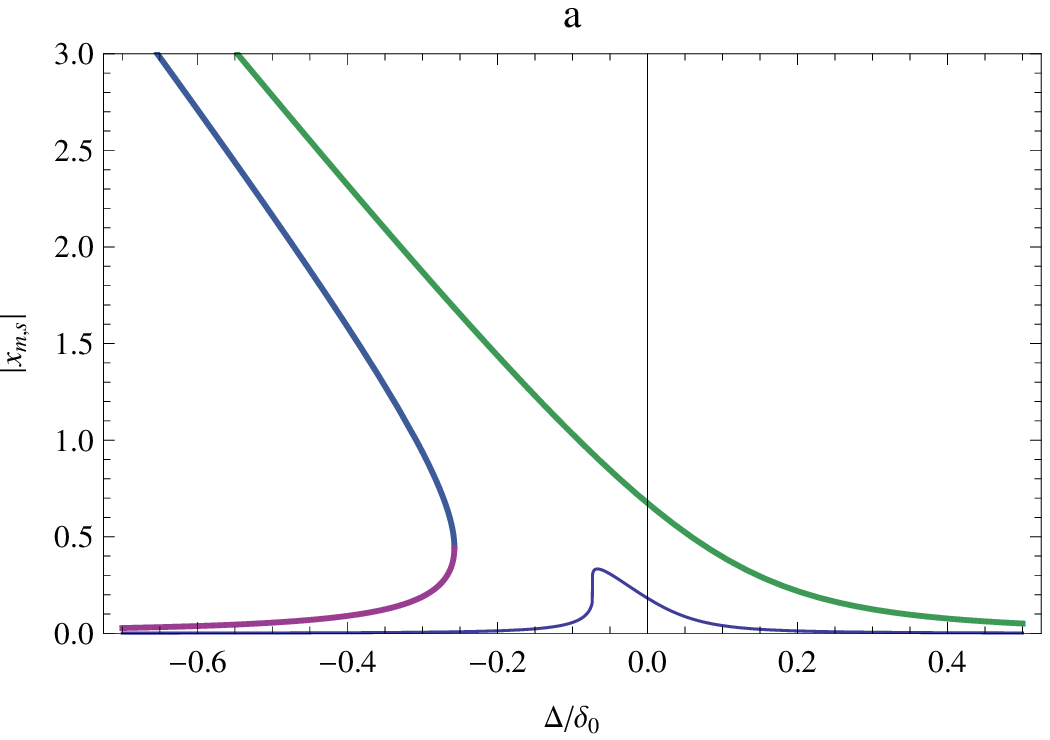}&\includegraphics[scale=0.7]{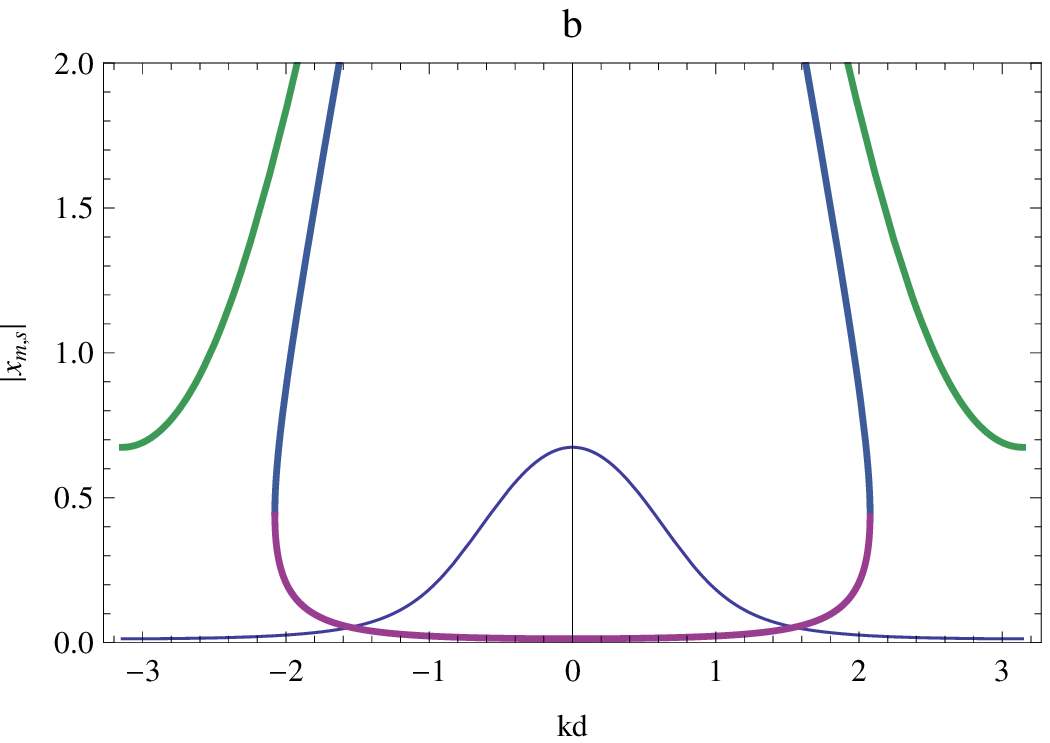}
\end{tabular}
\caption{(a): Steady state displacement spectrum as a function of $\Delta/\delta_{0}$ for $k=0$. The thin curve corresponds to pump intensity $\eta/\delta_{0}=0.2$ and the thick line corresponds to $\eta/\delta_{0}=1.0$. For the larger pump intensity, we find three steady-state solutions, with two of them being stable. The system prepared below the resonance will follow the steady state branch until reaching the lower turning point, where a non-steady state dynamics is excited. The parameters taken are, $\epsilon \Omega_{m}/\delta_{0}=0.2$, $\Omega_{m}/\delta_{0}=30$, $\Gamma_{m}/\delta_{0}=0.001$ and $\kappa/\delta_{0}=0.08$. (b): The displacement spectrum modified by the atomic backaction through the cavity photons as a function of the quasi-momentum ($kd$) for two different values of $\Delta_{c}/\delta_{0}$. For $\Delta_{c}/\delta_{0}=-0.5$, a bistable behaviour is seen (thick line) when the condensate is at the edge of the Brillouin zone ($kd=\pm \pi$ and $\Delta=0$). This bistable behaviour is absent when $\Delta_{c}/\delta_{0}=0.5$ (thin line).}
\label{fig.3}
\end{figure}

Here, $\Delta=\Delta_{c}-U_{0}N(J_{0}+2Jcos(kd))$. Note that now we do not ignore the tunneling term $J$. A plot of $|x_{m,s}|$ versus $\Delta/\delta_{0}$ for two different values of the pump parameter $\eta$ is shown in Fig. 3. Clearly for higher pump intensity the system shows a bistable behaviour. For pump rates higher than a critical value, we find three steady state solutions for the mirror displacement, with two of them being stable. The system prepared below resonance will follow the steady state branch until reaching the lower turning point, where a non-steady-state dynamics is excited. This dynamics is governed by the time scale of the mechanical motion of the mirror because the cavity damping is almost two orders of magnitude faster ($\Gamma_{m}<<\kappa$). Fig.3(b) shows the steady state displacement spectra  $|x_{m,s}|$ as a function of the quasi-momentum for two different values of the cavity detuning $\Delta_{c}/\delta_{0}=0.5$ (thin line) and $\Delta_{c}/\delta_{0}=-0.5$ (thick line). The cantilever phonons develop a quasi-momentum dependence due to strong coupling with the condensate mediated by the cavity photons. The atomic back-action modifies both the cavity field and the cantilever displacement. As the condensate moves across the Brillouin zone, the atom-field interaction changes and as a result the cantilever displacement spectra is continuously modified. For $\Delta_{c}/\delta_{0}=-0.5$, a bistable behaviour is seen when the condensate is at the edge of the Brillouin zone (here $\Delta=0$). This bistable behaviour is absent when $\Delta_{c}/\delta_{0}=0.5$. The position of the condensate in the Brillouin zone is easily manipulated by accelerating the condensate.

\section{Dynamics of small fluctuations: Normal Mode Splitting}

Here we show that the coupling of the mechanical oscillator, the cavity field fluctuations and the condensate fluctuations (Bogoliubov mode) leads to the splitting of the normal mode into three modes (Normal Mode Splitting(NMS)). The optomechanical NMS however involves driving three parametrically coupled nondegenerate modes out of equilibrium. The NMS does not appear in the steady state spectra but rather manifests itself in the fluctuation spectra of the mirror displacement. To this end, we shift the canonical variables to their steady-state values (i.e.~$\hat{a}\rightarrow \hat{a}_{s}+\hat{a}$ , $\hat{a}_{m}\rightarrow \hat{a}_{m,s}+\hat{a}_{m}$, $\hat{b}_{j}\rightarrow \dfrac{1}{\sqrt{M}}(\sqrt{N}+\hat{b})$) and linearize to obtain the following Heisenberg-Langevin equations neglecting atomic losses due to heating:

\begin{equation}\label{H-L0}
\dot{\hat{b}}=-i\left\lbrace \nu +2 U_{eff}\right\rbrace \hat{b}-iU_{eff}\hat{b}^{\dagger}-ig_{c}(\hat{a}+\hat{a}^{\dagger}) 
\end{equation}

\begin{equation}\label{H-L1}
\dot{\hat{a}} =\left(i\Delta_{d}-\frac{\kappa}{2}\right)\hat{a}-i\frac{g_{m}}{2}\left(\hat{a}_{m}+\hat{a}_{m}^{\dagger}\right)-ig_{c}(\hat{b}+\hat{b}^{\dagger})+\sqrt{\kappa}\xi_{p}(t),
\end{equation}

\begin{equation}\label{H-L2}
\dot{\hat{a}}_{m} =\left(-i\Omega_{m}-\frac{\Gamma_{m}}{2}\right)\hat{a}_{m}-i\frac{g_{m}}{2}\left(\hat{a}+\hat{a}^{\dagger}\right)+\sqrt{\Gamma_{m}}\xi_{m}(t).
\end{equation}

Here, $U_{eff}=\dfrac{Un_{0}}{\hbar}$, $g_{c}=U_{0}J_{0}\sqrt{N}|\hat{a}_{s}|$ , $\nu= U_{0}J_{0}|\hat{a}_{s}|^{2}+\dfrac{V_{cl}J_{0}}{\hbar}+\dfrac{E_{0}}{\hbar}$, $\Delta_{d}=\Delta_{c}-U_{0}NJ_{0}-\epsilon \Omega_{m} x_{m,s}$ is the detuning with respect to the renormalized resonance and $\Delta_{d}<0$ leads to cooling. Note that $x_{m,s}$ is negative and hence makes $\Delta_{d}$ positive. The atomic and cantilever back-action modifies the cavity detuning.  The optomechanical coupling rate is given by $g_{m}=2\hat{a}_{s}\epsilon \Omega_{m}$ and  $|\hat{a}_{s}|^{2}$ gives the mean
resonator occupation number. In deriving the above equation, we have assumed that $\hat{a}_{s}$ to be real. As before, we assume negligible tunneling ($J=E=0$) and hence we drop the site index $j$ from the atomic operators.. We will always assume $\Gamma_{m}\ll\kappa$. Equations~(\ref{H-L0},\ref{H-L1},\ref{H-L2}) and their Hermitian conjugates constitute a system of six first-order coupled operator equations, for which the Routh-Hurwitz criterion implies that the system is stable only for $g_{m}<\sqrt{(\Delta_{d}\Delta_{d}^{'}+\kappa^{2}/4)\Omega_{m}/|\Delta_{d}|}$ and $\Delta_{d}\Delta_{d}^{'}>0$, where $\Delta_{d}^{'}=\Delta_{d}-2g_{c}/(\nu+3 U_{eff})$. We transform to the quadratures: $X_{m}=\hat{a}_{m}+\hat{a}_{m}^{\dagger}$, $P_{m}=i(\hat{a}_{m}^{\dagger}-\hat{a}_{m})$,  $X_{p}=\hat{a}+\hat{a}^{\dagger}$, $P_{p}=i(\hat{a}^{\dagger}-\hat{a})$, $X_{b}=\hat{b}+\hat{b}^{\dagger}$, $P_{b}=i(\hat{b}^{\dagger}-\hat{b})$. The displacement spectrum in Fourier space for $n_{p}=0$ is found as

\begin{equation}\label{displacement_spectra}
S_{x}(\omega)=\frac{x_{0}^{2}}{2\pi}\Omega_{m}^{2}|\chi(\omega)|^{2}\left[\Gamma_{m}n_{m}-\frac{\Delta_{d}^{2}+\omega^{2}+\kappa^{2}/4}{2\Delta_{d}\Omega_{m}}\Gamma_{s}(\omega)\right],
\end{equation}

where,

\begin{equation}
\chi^{-1}(\omega) =\Omega_{m}^{2}+2\Omega_{m}\Omega_{s}(\omega)-\omega^{2}-i\omega\left[\Gamma_{m}+\Gamma_{s}(\omega)\right]
\end{equation}

\begin{equation}
\Omega_{s}(\omega) = \dfrac{\Delta_{d} g_{m}^{2} \Upsilon^{4} \Omega_{\nu}^{2} }{2 \left\lbrace \Upsilon^{8}+\kappa^{2} \omega^{2}\Omega_{\nu}^{4}\right\rbrace }
\end{equation}

and 

\begin{equation}
\Gamma_{s}(\omega) = -\dfrac{ \kappa \Delta_{d} g_{m}^{2} \Omega_{m}\Omega_{\nu}^{4} }{ \left\lbrace \Upsilon^{8}+\kappa^{2} \omega^{2}\Omega_{\nu}^{4}\right\rbrace }
\end{equation}

\begin{equation}
\Upsilon^{4}=\left\lbrace (\dfrac{\kappa^{2}}{4}+\Delta_{d}^{2}-\omega^{2})(\omega^{2}-(\nu+U_{eff})(\nu+3U_{eff}))-4\Delta_{d}g_{c}^{2}(\nu+U_{eff})\right\rbrace
\end{equation}

\begin{equation}
\Omega_{\nu}^{2}= \left\lbrace \omega^{2}-(\nu+U_{eff})(\nu+3U_{eff})\right\rbrace 
\end{equation}

This spectrum is characterized by a mechanical susceptibility $\chi(\omega)$ that is driven by thermal noise ($\propto n_{m}$) and by the quantum fluctuations of the radiation pressure (quantum back-action) and quantum fluctuations of the condensate. Note that in the absence of the atom-photon coupling ($g_{c}=0$), the displacement spectrum Eqn.\ref{displacement_spectra} reduces to that found in \cite{dobrindt}.

\begin{figure}[t]
\begin{tabular}{cc}
\includegraphics[scale=0.7]{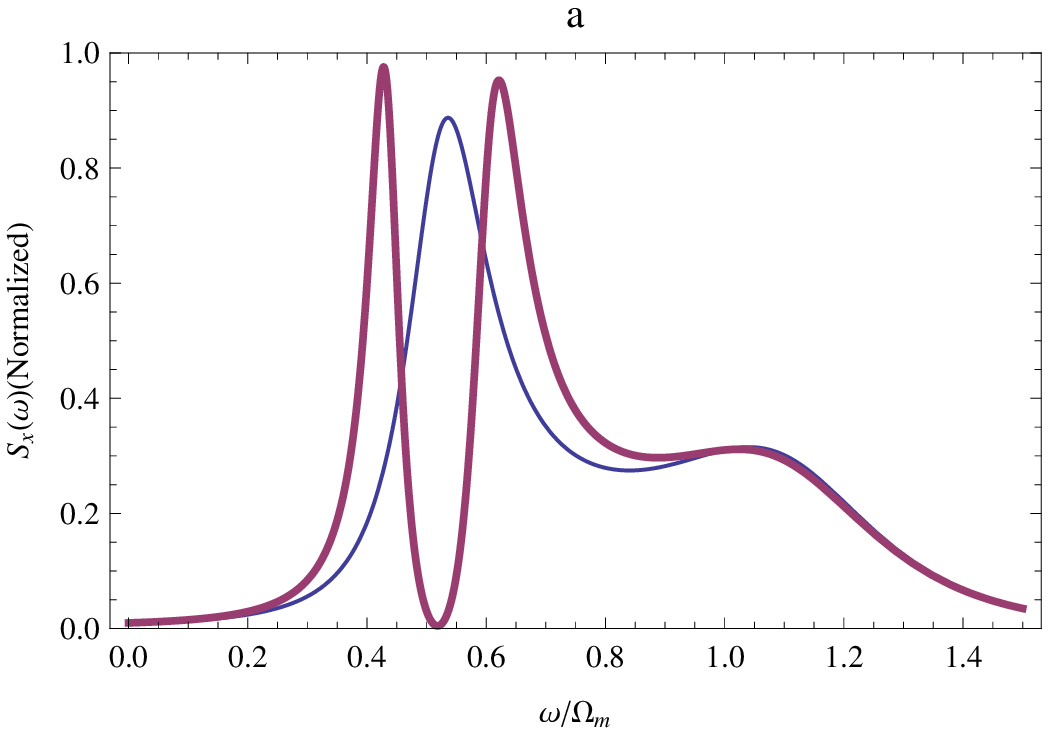}& \includegraphics[scale=0.7]{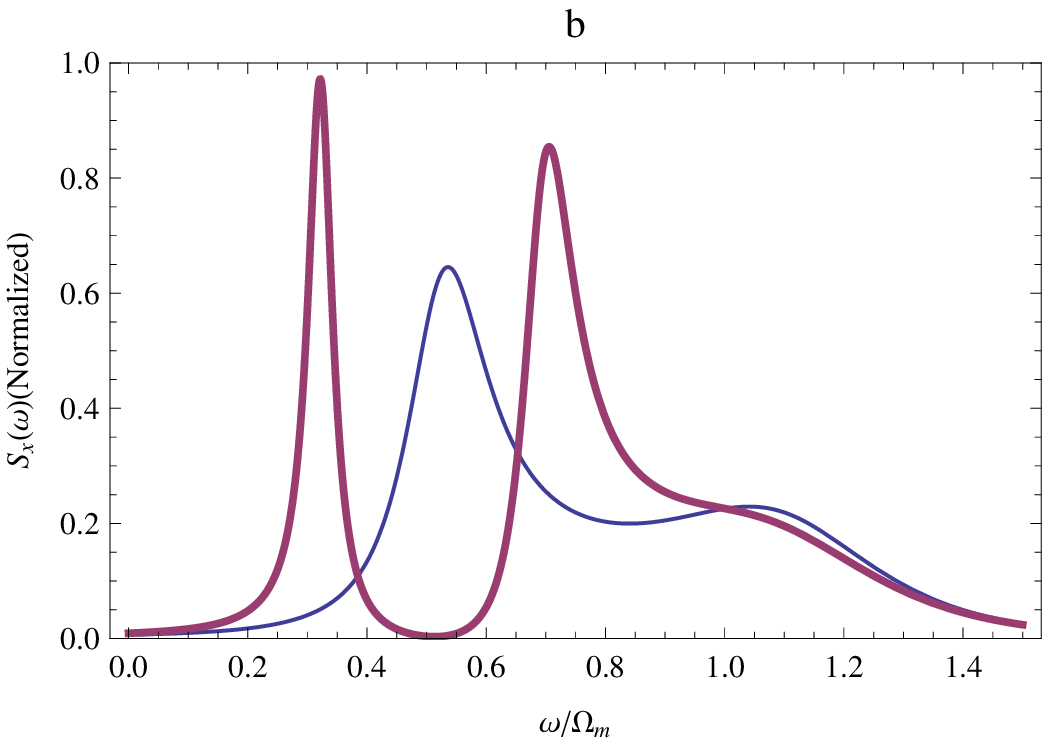}
\end{tabular}
\caption{(a): Normalized plot of the displacement spectrum $S_{x}(\omega)$ for two values of the atomic two-body interaction, $U_{eff}/\Omega_{m}=0$ (thin line) and $U_{eff}/\Omega_{m}=0.3$ (thick line). Parameters chosen are: $g_{m}/\Omega_{m}=0.4$, $\kappa/\Omega_{m}=0.1$, $\Gamma_{m}/\Omega_{m}=0.01$, $g_{c}/\Omega_{m}=0.1$, $\Delta_{d}/\Omega_{m}=-1$ and $\nu/\Omega_{m}=0.01$. A clear NMS is observed in the presence of atom-atom interaction. , (b): Normalized plot of the displacement spectrum $S_{x}(\omega)$ for the same parameters as (a) but with a stronger atom-photon coupling $g_{c}/\Omega_{m}=0.2$. The NMS is now much more prominent compared to that in (a). }
\label{fig.4}
\end{figure}

In Fig. 4a, we show a normalized plot of the displacement spectrum $S_{x}(\omega)$ for two values of the atomic two-body interaction, $U_{eff}/\Omega_{m}=0$ (thin line) and $U_{eff}/\Omega_{m}=0.3$ (thick line). Parameters chosen are: $g_{m}/\Omega_{m}=0.4$,$\kappa/\Omega_{m}=0.1$,$\Gamma_{m}/\Omega_{m}=0.01$, $g_{c}/\Omega_{m}=0.1$, $\Delta_{d}/\Omega_{m}=-1$ and $\nu/\Omega_{m}=0.01$. In the absence of the interactions, we observe the usual normal mode spliiting into two modes and we find that due to a finite atom-atom interaction the normal mode splits up into three modes.  The NMS is associated with the mixing between the mechanical mode and the fluctuation of the cavity field around the steady state and the fluctuations of the condensate (Bogoliubov mode) around the mean field. The origin of the fluctuations of the cavity field is the beat of the pump photons with the photons scattered from the condensate atoms. The frequency of the Bogoliubov mode in the low momentum limit is $\approx$ $\sqrt{U_{eff}}$. Hence in the absence of interactions, the Bogoliubov mode is absent and the system simply reduces to the case of two mode coupling (i.e coupling between the mechanical mode and the photon fluctuations). In the presence of finite atom-atom interaction, the mechanical mode, the photon mode and the Bogoliubov mode forms a system of three coupled oscillators. It is important to note that splitting in the displacement spectrum occurs only for $g_{m}>\kappa/\sqrt{2}$ due to finite width of the peaks. Fig. 4(b) illustrates the influence of increasing the atom-photon coupling $g_{c}=0.2$. The NMS is now much more prominent compared to the case with $g_{c}=0.1$. Similar three coupled oscillator experimental results where two coupled cavities, each containing three identical quantum wells \cite{stanley} and one micro-cavity containing two quantum wells\cite{lindmark} have been reported. One experimental limitation could be spontaneous emission which leads to momentum diffusion and hence heating of the atomic sample \cite{murch}. The parameters chosen here are within current technological reach\cite{schliesser2,murch,brennecke} and hence the observation of the paramertic NMS is within experimental reach.

\section{Conclusions}

In summary we have analyzed a novel scheme of cavity-opto-mechanics with ultracold atoms. We showed that the steady state displacement spectra of a cantilever coupled indirectly to a gas of ultracold atoms in an optical lattice through the cavity field are distinct for different quantum phases of equal densities. Further, we showed that in the mean-field of the atoms, the cantilever displacement shows a bistable behviour for high pump intensities and exhibits a dependence on the position of the condensate in the Brillouin zone due to the back-action of the condensate on the cantilever. The cantilever displacement spectrum shows a bistable behaviour when the condensate is at the edge of the Brillouin zone. In the presence of atom-atom interactions, the coupling of the mechanical oscillator, the cavity field fluctuations and the condensate fluctuations (Bogoliubov mode) leads to the splitting of the normal mode into three modes (Normal Mode Splitting).  The system described here shows a complex interplay between distinctly three systems namely, the nano-mechanical cantilever, optical micro-cavity and the gas of ultracold atoms. The quantum state of the degenerate gas influences the cantilever displacement spectra via the cavity photons, while on the other hand the cantilever displacement modifies the cavity field which in turn modifies the properties of the condensate. Manipulating the position of the condensate in the Brillouin zone and the atomic tunneling, we now have a new possibility to efficiently control the cantilever dynamics and at the same time the cantilever dynamics can be used to control the atomic dynamics.

\end{document}